\documentclass[a4paper,tikz]{article}

\usepackage{bm}
\usepackage[top=1in,left=1in,bottom=1in,right=1in]{geometry}
\usepackage{amsmath, amssymb, mathtools}
\usepackage{braket}
\usepackage{tikz}
\usetikzlibrary{arrows.meta, decorations.pathmorphing}
\begin{document}

\title{On the dynamical evolution of randomness\\ Part B: Geometrisation and the origin of convergence in LLN }

\author{Allen Lobo\thanks{allen.e.lobo@outlook.com}\\Dept. of Physics, St. Joseph's University, Bengaluru, Karnataka, India.}
\maketitle

\begin{abstract}
In classical probability theory, the convergence of empirical frequencies to theoretical probabilities — as captured by the Law of Large Numbers (LLN) — is treated as axiomatic and emergent from statistical assumptions such as independence and identical distribution. In this work, a novel dynamical framework is constructed in which convergence arises as a consequence of structured evolution in outcome space, rather than a statistical postulate.  

Through this formalism, statistical convergence is derived dynamically, revealing an internal structure to randomness and exposing entanglement between successive trials. The system recovers classical LLN behaviour in the large-number limit, while predicting deviations, transient fluctuations, and geometric asymmetries in the early regime. This work inaugurates a new paradigm — \textit{dynamical probability mechanics} — in which randomness is modelled not as a sequence of disconnected stochastic events, but as a physically structured, feedback-driven process. The theory provides a novel explanatory layer beneath statistical laws and opens pathways toward a mechanistic foundation of probability itself.
\end{abstract}

\section{Introduction}
The Law of Large Numbers (LLN) \cite{Bolthausen2013BernoullisNumbers} asserts that the relative frequencies of outcomes in repeated random experiments converge to fixed theoretical probabilities as the number of trials increases. This statistical convergence is one of the most deeply accepted principles in probability theory and is foundational to statistical physics, machine learning, and experimental science. Yet, despite its ubiquity, LLN is rarely questioned in terms of its physical or dynamical origin. In prevailing theory, randomness is treated as ontologically primitive: outcomes are assumed to be independent and identically distributed (i.i.d.), and convergence is derived mathematically from these axioms. While this treatment is mathematically complete within the Kolmogorov's axiomatic framework \cite{Kolmogorov1933FoundationsProbability}, it offers no insight into why empirical frequencies stabilize over time. In other words, the convergence itself is a law — but not one with an underlying physical mechanism.

Let an ideal random event  $\bm{\Psi}$  have $n$ outcomes, each represented by $\psi_i$, where $i\in [1,n]$. By ideal it is stressed that each outcome is i.i.d. In this regard, the measure of probability of each outcome, when the event occurs with one and only one outcome, is 
$$ P_i = \frac{1}{n},$$
this is well known. Suppose this event is repeated for $m$ number of times; the measure of how many times an outcome $i$ occurs ($m_i$), relative to the total number of outcomes, can be measured as:
\begin{equation}\label{emp_prob}
    \frac{m_i}{m} = \mathcal{L}_i.
\end{equation}
This measure is the empirical probability of the outcome. Equation (\ref{emp_prob}) states that 
\begin{equation}
    \sum_{i=1}^n\mathcal{L}_i = 1.
\end{equation}
The dynamical behaviour of such a random event, occurring in repetition, describes the evolution of $\mathcal{L}_i$ associated with each outcome $i\in[1,n]$ of the random event. It is possible (however unlikely) for a single outcome ($k$) to occur each time, such that $\mathcal{L}_i=\delta_{ki}$, where $\delta_{ki}$ is the Dirac-delta function, and is 1 for $i=k$, zero otherwise. However, this does not happen. From Poincaré's Recurrence Theorem (and the Ergodic principle), it is understood that in a system of finite, countable possible outcomes, over large repetitions of the experiment, each outcome would be observed multiple times. When the number of repetitions become close to infinity, the number of recurrences of each outcome become equal; this is precisely the conclusion obtained from the LLN. But these theories do not describe the hidden dynamics of statistics which causes each outcome to occur equally in number, as the number of trials reaches infinity.

It is known from Bernoulli's LLN \cite{Chibisov2016BernoullisNumbers, Mattmuller2014The1713} that over continuous repetitions, each equally probable outcome of a random experiment becomes empirically equally likely, such that
\begin{equation}
\lim_{j\to \infty}\, ^j\mathcal{L}_i \rightarrow P_i.
\end{equation}
Here, the superscript $j$ denotes the number of trials performed so far. Although mathematical proof of LLN, along with its applications and further development, has been presented by many authors \cite{Chibisov2016BernoullisNumbers, Teran2008OnNumbers, Weba2009AMethod, Goldstein1975SomeNumbers, Dedecker2007ApplicationsNumbers, Teran2006AApplications, SHIRIKYAN2003AAPPLICATIONS, Yang2008AApplications, Kay2014LawsApplications}, its physical explanation remains unanswered. It is obvious that the LLN works: the probability that each outcome occurs equally frequent becomes
\begin{gather}
   P(m_i=m_k \,\forall i,k \in [1,n])= \frac{m!}{\left(\frac{m}{n}!\right)^n}\cdot\left(\frac{1}{n}\right)^m,
    \intertext{which is the maximum of the general case,}
    P=\frac{m!}{m_1!\cdot m_2!\cdot m_3! ...m_n!}\cdot\left(\frac{1}{n}\right)^m. 
\end{gather}
Here $m_a$ is the frequency of outcome $a$. Any extrema expression of the outputs, such as one output coming always and other outputs occurring never, lie on extremely low values of the probability distribution of output states. Hence, the most probable outcome is the one predicted by the LLN, and is even further enhanced by increasing $m$ to $\infty$. Also note that this limit gives the same result as would be the case for infinite possible outcomes, since, in the limit
\begin{equation}
    \lim_{n\to\infty}\left(\frac{1}{n}\right)^m \approx \left(\frac{1}{m}\right)^n , \quad\forall \,m,n>0. 
\end{equation}

In a very recent work \cite{Lobo2025OnTheorem}, a peculiarly intuitive outcome of a random experiment occurring in repetition has been demonstrated: outcomes with less frequent occurrences, in a repeated random event, gain their frequencies, and outcomes which initially occur more frequently, reduce the frequency of their occurrence. This occurs such that as the number of trials increase, each outcome gains a nearly equal frequency of occurrence, as suggested by the LLN. This is expressed in the relation \cite{Lobo2025OnTheorem}: 
\begin{equation}\label{dmkdn}
     \frac{dm_k}{dm} = m_k\frac{d}{dm}\ln{\Lambda_\sigma} + \mathcal{L}_k n - \sum_{k\neq i}^{n}\frac{\mathcal{L}_k}{\mathcal{L}_i}\frac{dm_i}{dm}.
\end{equation}
The modulation in the growth-rate of outcomes $m_k$ is presented by the term involving the ratio ${\mathcal{L}_k}/{\mathcal{L}_i}$. As the frequency $(m_k)$ of an outcome $\psi_k$ decreases, the ratio invariably decreases and therefore the growth rate $d_mm_k$ increases. This suggests some entanglement, some coupling between the occurrences of each outcome of a random event.

In this work, I aim to explore this entanglement (in successive repetitions) between completely independent outcomes of an ideal random event. By independence, it is implied that each outcome is independent of the other, at at a certain trial of the random experiment. This means that for each $^j\bm{\Psi}$, 
$$^j\braket{\psi_i|\psi_k}=\delta_{ik}.$$ 
The entanglement is, in this work, thought to exist in successive trials. With repetition, convergence is found to exist in a quantity termed the $\Lambda-$entropy $(\Lambda_\sigma)$ \cite{Lobo2025OnTheorem}, which is equal to the product of all $\mathcal{L}$ up-to any given repetitions $j$,
\begin{gather}
    ^j\Lambda_\sigma = \prod_{i=1}^n\,^j\mathcal{L}_i = \frac{1}{M^n}\prod_{i=1}^n\,^jm_i.
    \intertext{The dynamics of this entropy term is given as \cite{Lobo2025OnTheorem}:}
    \frac{d}{dm}\Lambda_\sigma = \Lambda_\sigma \left[ \frac{d }{dm}\ln \left(\prod_{i=1}^n m_i\right) -\frac{n}{m} \right].\label{dlambdadm}
    \intertext{From the Law of Large Numbers, one gets}
    \lim_{j\to\infty}\,^j\Lambda_\sigma = \prod_{i=1}^n P_i = \Lambda_\sigma^0.
\end{gather}
By \textit{dynamics}, it is meant that the evolution of $\mathcal{L}_i$ and $\Lambda_\sigma$, along-with the entanglement of $^j\mathcal{L}_i$ with $^j\mathcal{L}_k$, where $i,k \in [1,n]$, is observed with $j\in[0,M]$, instead of time. Though this evolution is discrete, with $\Delta j = 1$ at each step, the model may be perceived as continuous, typically when dealing with temporally repetitive random systems or systems with large number of repetitions. At large, the deterministic nature of ideally random events, as also displayed by nature in experiments such as the Monte-Carlo setups, should be revealed by exploring these dynamics. 

 \section{Dynamics of random events}
As presented initially by Lobo and Arumugam \cite{Lobo2025OnTheorem}, the convergence of empirical randomness describes dynamical equations which can be used to portrait the evolution of the random experiment. We consider a random experiment with $n$ outcomes, each having probability $P_i$ equal to $1/n$, $i\in [1,m]$. We repeat the experiment $m$ number of times, and measure the frequency of each outcome as $m_i$. The empirical probability $\mathcal{L}_i$ is then calculated as the ratio of the outcome frequency to total number of repetitions, 
\begin{equation}
    \mathcal{L}_i = \frac{m_i}{n}.
\end{equation}

A new measure of the "randomness" of a system has been presented by Lobo and Arumugam \cite{Lobo2025OnTheorem} as the $\Lambda-$entropy $(\Lambda_\sigma)$. The LLN dictates that the empirical probabilities must converge to the theoretical probabilities as $m\rightarrow\infty$. The $\Lambda-$entropy must also converge to $\Lambda_\sigma^0$, where,
\begin{equation}\label{lambda_sigma_eqn}
    \Lambda_\sigma^0 = \prod_{i=1}^m P_i.
\end{equation}
Hence, two limits can be presented to the system,
    \begin{gather}
         \lim_{m\to\infty}\Lambda_\sigma = \Lambda_\sigma^0,\label{axiom1.1}\\
         \lim_{m\to\infty}\frac{d}{dn}\Lambda_\sigma =0.\label{axiom1.2}
    \end{gather}

As the experiment starts and proceeds with repetitions such that $m\geq n$, $\Lambda_\sigma$ proceeds away from $\Lambda_\sigma^0$. This occurs not due to bias in the outcomes but due to \textit{chance}. The empirical randomness of the experiment can be measured by measuring the relative shift of $\Lambda_\sigma$ from the $\Lambda_\sigma^0$ value. This approach provides a measure of the empirical randomness of a random experiment. Alternatively speaking, if an experiment with well-defined outcomes is more random, as the experiment proceeds into repetitions, the $\Lambda-$entropy remains closer to its ideal value $\Lambda_\sigma^0$. The greater $\Lambda_\sigma$ shifts from the $\Lambda_\sigma^0$ value, lesser is the empirical randomness of the experiment. i.e.,
    \begin{equation}
       \text{Empirical Randomness} \propto \frac{1}{|\Lambda_\sigma - \Lambda_\sigma^0|}.\label{theorem1.1}
    \end{equation}
    This deviation of the empirical randomness as $\Lambda_\sigma$ deviates from $\Lambda_\sigma^0$ can be measured using a parameter $\sigma$, where,
    \begin{gather}
        \sigma = |\Lambda_\sigma - \Lambda_\sigma^0|,
        \intertext{such that,}
        \text{Empirical Randomness} \propto \frac{1}{\sigma}.
    \end{gather}

Empirical randomness increases as the empirical randomness parameter $(\sigma)$ approaches zero. Upon attempting to investigate how the $\Lambda-$entropy varies with the trials of the experiment, we find its derivatives with respect to $m$.

Differentiating $\Lambda_\sigma$ with respect to $n$, we get --
\begin{gather}\allowdisplaybreaks
    \frac{d\Lambda_\sigma}{dm} = \frac{d}{dm}\left(\frac{1}{m^n}\prod_{i=1}^{n}m_i\right) \\
    = \left(\prod_{i=1}^{n}m_i\right)\frac{d}{dm}\frac{1}{m^n} + \frac{1}{m^n}\left(\frac{d}{dm}\prod_{i=1}^{n}m_i\right).\\
    = -\frac{n}{m}\frac{1}{m^n}\left(\prod_{i=1}^{n}m_i\right) + \frac{1}{m^n}\sum_{i=1}^n\left(\prod_{j=1}^{n}f_j\right)\frac{1}{m_i}\frac{dm_i}{dm}.\\
   \therefore \frac{d}{dm}\Lambda_\sigma = \Lambda_\sigma \left( -\frac{n}{m} + \sum_{i=1}^n \frac{d }{dm}\ln m_i \right). 
   \intertext{The summation in the above equation can be further reduced to simplify the form, giving --}
   \frac{d}{dm}\Lambda_\sigma = \Lambda_\sigma \left[ \frac{d }{dm}\ln \left(\prod_{i=1}^n m_i\right) -\frac{n}{m} \right].\label{derivative_of_lambda_entropy_first_principle}
\end{gather}

It can be seen from equation (\ref{derivative_of_lambda_entropy_first_principle}) that the number of outcomes $(n)$ reduces the convergence rate of $\Lambda-$entropy towards the $\Lambda_\sigma^0$ value. It may be noted that in the case where $m<n$, at-least one outcome has not yet occurred, such that both $\prod_{i=1}^n m_i$ and $\Lambda_\sigma$ remain zero. Hence, $\Lambda_\sigma$ gains magnitude only when all the outcomes have occurred at-least once. Alternatively, for the case when $m\rightarrow\infty$, it can be shown that $\Lambda_\sigma$ remains zero until the number of trials itself becomes large enough, such that $\Lambda_\sigma$ becomes equal to $\Lambda_\sigma^0$.

For small values of $m$, provided the condition that $m\geq n$, simple manipulation of equation (\ref{derivative_of_lambda_entropy_first_principle}) presents the nature of flow of outcomes during the repetition of the experiment. From equation (\ref{derivative_of_lambda_entropy_first_principle}) we have,
\begin{gather}
    \frac{1}{\Lambda_\sigma}\frac{d}{dm}\Lambda_\sigma  +\frac{n}{m}= \frac{d }{dm}\ln \left(\prod_{i=1}^n m_i\right) = \left(\prod_{i=1}^n m_i^{-1}\right)  \frac{d }{dm}\prod_{i=1}^n m_i\\
    \Rightarrow  \frac{d }{dm}\prod_{i=1}^n m_i=\left(\prod_{i=1}^n m_i\right)\cdot\left[\frac{d}{dm}\left(\ln\Lambda_\sigma\right)  +\frac{n}{m} \right].
    \intertext{Expressing the above equation in terms of the empirical randomness parameter $(\sigma)$,}
    \begin{split}
    \frac{d }{dm}\prod_{i=1}^n m_i=\left(\prod_{i=1}^n m_i\right)&\cdot\left[\frac{d}{dm}\ln\left(\Lambda_\sigma^0 + \delta\sigma\right)  +\frac{n}{m} \right],\\
    \text{where,}\quad\delta &= \begin{cases}
        +1, & \Lambda_\sigma > \Lambda_\sigma^0\\
        -1, & \Lambda_\sigma < \Lambda_\sigma^0\\
        0, & \Lambda_\sigma = \Lambda_\sigma^0.
    \end{cases}.
    \end{split}\label{chance_equation}
\end{gather}

Equation (\ref{chance_equation}) states that the rate at which outcomes of an experiment occur such that each outcome gains its frequency of occurrence as the number of trials increase, depend directly on how empirically random the experiment is at its current stage, which is measured by $\sigma$. In other words, if the random experiment proceeds in such a way that the frequency of outcomes deviate from uniformity, the empirical randomness parameter $(\sigma)$ increases, due to which the rate at which $\prod_{i=1}^n m_i$ grows per trial also increases, which implies that outcomes with currently less occurrence frequencies $m_i$ also increase such that $\prod_{i=1}^n m_i$ grows more rapidly. Outcomes of a random experiment are therefore not merely governed by \textit{chance}, but depend on the instantaneous empirical randomness of the experiment as it is repeated. Indeed as $n\rightarrow \infty$, equation (\ref{chance_equation}) dictates that the product $\prod_{i=1}^m$ reaches a maxima.

In order to determine the evolution of the frequencies of each outcome, we initially take the case of an ideal experiment with two outcomes $(n=2)$, such as the coin-toss experiment. From equation (\ref{derivative_of_lambda_entropy_first_principle}), we get:
\begin{gather}
    \frac{d}{dm}\ln{\Lambda_\sigma} + \frac{n}{m} = \sum_{i=1}^{n} \frac{1}{m_i}\frac{d}{dm}m_i = \frac{1}{m_H}\frac{dm_H}{dm}+\frac{1}{m_T}\frac{dm_T}{dm},
    \intertext{where $m_H$ and $m_T$ represent the frequencies of occurrence of Head and Tail as outcomes of the coin toss, respectively. Multiplying both sides by $m$ gives:}
    m\frac{d}{dm}\ln{\Lambda_\sigma} + n =  \frac{m}{m_H}\frac{dm_H}{dm}+\frac{m}{m_T}\frac{dm_T}{dm}=\frac{1}{\mathcal{L}_H}\frac{dm_H}{dm}+\frac{1}{\mathcal{L}_T}\frac{dm_T}{dm}\\
    \Rightarrow \frac{dm_H}{dm} = -\frac{\mathcal{L}_H}{\mathcal{L}_T}\frac{dm_T}{dm} + n\mathcal{L}_H + m\mathcal{L}_H\frac{d}{dm}\ln{\Lambda_\sigma}\\
    \Rightarrow \frac{dm_H}{dm} = -\frac{\mathcal{L}_H}{\mathcal{L}_T}\frac{dm_T}{dm} + n\mathcal{L}_H + m_H\frac{d}{dm}\ln{\Lambda_\sigma}.
\end{gather}
 The above expression can be now generalised as follows:
 \begin{gather}
     \frac{dm_k}{dm} = m_k\frac{d}{dm}\ln{\Lambda_\sigma} + n\mathcal{L}_k  - \sum_{k\neq i}^{n}\frac{\mathcal{L}_k}{\mathcal{L}_i}\frac{dm_i}{dm},\label{gen_eqn1}
 \end{gather}
 From equation (\ref{gen_eqn1}), the original equation (\ref{derivative_of_lambda_entropy_first_principle}) can be extracted by multiplying with $m_i^{-1}$ on both sides. Here, one must note that the equation holds validity only after each outcome has occurred at-least once, such that $\mathcal{L}_i>0$, to avoid singularity.

 In the limit $n\rightarrow\infty$, the following adjustments can be made, as previously discussed:
 \begin{gather}
     \lim_{m\to\infty}\begin{cases}
         \Lambda_\sigma \rightarrow\Lambda_\sigma^0,\\
         \frac{d}{dm}\ln\Lambda_\sigma \rightarrow 0,\\
         \mathcal{L}_i \rightarrow P_i=\frac{1}{n},\\
         \therefore\sum_{k\neq i}^{n}\frac{\mathcal{L}_k}{\mathcal{L}_i}\frac{dm_i}{dm} = \sum_{k\neq i}^{n}\frac{dm_i}{dm}= \frac{d \sum_{k\neq i}^{n}m_i}{dm} = 1,\\
         \text{and, }m_{i,j} \rightarrow\infty.
     \end{cases}
     \intertext{This modifies equation (\ref{gen_eqn1}) as follows:}
     \lim_{m\to \infty} \frac{dm_j}{dm} = \mathcal{E} + 1 - 1 = \mathcal{E},\label{uncertainity_inlimit}
 \end{gather}
 where,  $\mathcal{E} = \lim_{m\to\infty}m_k\frac{d}{dm}\ln{\Lambda_\sigma} = \infty\times 0$ is the indeterminate term. What is interesting is that in this limit, the rate of growth of each outcome becomes equal to $\mathcal{E}$, independent of the growth rate of other outcomes, as is seen in the case of small values of $m$. The value of $\mathcal{E}$ can, however, only belong to the finite set $\{0,1\}$, since at each trial, the outcome frequencies can increase either by $1$ or remain same\footnote{This is valid even for the case of finite values of $m$}. Also, only one outcome can have $\mathcal{E}=1$ at each trial, with the others having growth rates as $0$. The indeterminate form in this case arises by the inability of prediction of the growth rate of the outcomes based on the growths of other outcomes, as seen in equations (\ref{gen_eqn1}) and (\ref{gen_eqn2}). This means that in the limit $n\rightarrow \infty$, the outcomes of the experiment become truly indeterminable. They do not follow the probability evolution in the form of equations (\ref{gen_eqn1}) or (\ref{derivative_om_lambda_entropy_first_principle}), and therefore can not be predicted, such as in equation (\ref{integration}). Therefore, the outcomes achieve a true state of \textit{randomness}, which the theoretical model predicts. It therefore makes sense that in such a state, the experiment displays natures of ideal randomness, which is confirmed by the divergence of $1/\sigma$ to infinity and convergence of $\Lambda_\sigma$ to $\Lambda_\sigma^0$. This also implies that as the experiment proceeds, the term $m_i-m_j$ must obtain truly random values, since the result would depend on the integration of an indeterminable term:
 \begin{gather}\label{ft-fh}
   \forall i\neq k, \quad \lim_{m\to\infty}  m_i - m_k = \int\mathcal{E}dn  - m_k.
 \end{gather}
  
What is obvious from these results is that the actual dynamical evolution of the empirical probabilities of the outcomes of an experiment, and the associated measures of $\Lambda_\sigma$, are independent of the theoretical probability predictions $(P_i)$. As a matter of fact, empirical probabilities are completely independent of their theoretical counterparts, besides the fact that the point of convergence is itself the theoretical probabilities. Hence, it appears that the influence of empirical randomness and associated convergence is an orthogonal phenomena to the probability descriptions of the experiment. In the next section, the act of the experiment itself, and its associated outcomes, are given a vector structure. The act of the experiment and its repetitions is shown to be a vector operation in an outcome space, and corresponding results are presented. 
 
\section{Vector modelling of the random experiment}
Using a vector approach, the orthogonal outcomes (suppose, three) of a given  $\bm{\Psi}$  could be represented as vector states $\ket{\psi_i}$ in terms of basis vectors $\ket{\phi}$,
\begin{gather}
 \ket{\phi_1}=\begin{pmatrix}
        1\\0\\0
    \end{pmatrix}, \ket{\phi_2}=\begin{pmatrix}
        0\\1\\0
    \end{pmatrix},\ket{\phi_3}=\begin{pmatrix}
        0\\0\\1
    \end{pmatrix}.\\
    \ket{\psi_1}=\sqrt{P_1}\begin{pmatrix}
        1\\0\\0
    \end{pmatrix}, \ket{\psi_2}=\sqrt{P_2}\begin{pmatrix}
        0\\1\\0
    \end{pmatrix},\ket{\psi_3}=\sqrt{P_3}\begin{pmatrix}
        0\\0\\1
    \end{pmatrix}.
    \intertext{It is obvious that the number of components of each state should be equal to $n$, the total number of possible outcomes. The probability (operator) $\hat P$ of a state can be given as}
    \hat{P}\ket{\psi_i} = \sum_{l=1}^n\frac{\bra{\psi_l}}{\sum_{k=1}^n\braket{\psi_k|\psi_l}}\ket{\psi_i},   
\end{gather}
which is simply the normalized inner product of the state with itself (if no entanglement exists between outcomes). The experiment  $\bm{\Psi}$  then represents a vector of  $n$ orthogonal components. 
\begin{gather}
    \ket{\bm{\Psi}} = \sum_{i=1}^n\ket{\psi_i} = \sum_{i=1}^n\sqrt{P_i}\ket{\phi_i}. 
    \intertext{For the case of the three-outcome state shown in equation (11), probability for each independent state (outcome) becomes $P_i = 1/3$, $\forall i =\{1,2,3\}$.  $\bm{\Psi}$  becomes}
    \bm{\Psi} = \ket{\psi_1} + \ket{\psi_2}+\ket{\psi_3} = \begin{pmatrix}
        1/\sqrt{3}\\\\1/\sqrt{3}\\\\1/\sqrt{3}
    \end{pmatrix}.
    \end{gather}

    \subsection*{Mechanical analogue of performing an experiment}
    Based on the vector model of the experiment, the problem of experimentation followed by presentation of one (true) outcome of the experiment, as well as its repetition, can be described as a mechanical or geometric transformation of the state vector. 
    The act of performing an experiment  $\bm{\Psi}$  , drawing one of possible outcomes $\psi_i$, can be modelled as rotating the vector  $\bm{\Psi}$  such that one (and only one) component (vector state) relates with an \textit{axis of reality} $(\mathcal{R_A})$. The axis of reality is a curve which depicts the true outcomes of experiments, occurring sequentially, by coinciding with the outcome vector $\phi$ after the experiment is performed. This coincidence determines the true outcome of the experiment, which otherwise consists of $n$ possibilities. This axis must exist within the dimensions of  $\bm{\Psi}$  itself, since at-least one outcome must always occur. The axis of reality can be interpreted as an axis which \textbf{records} the events occurring out of an experiment.
     \begin{figure}[!ht]
        \centering
        \includegraphics[width=1\linewidth]{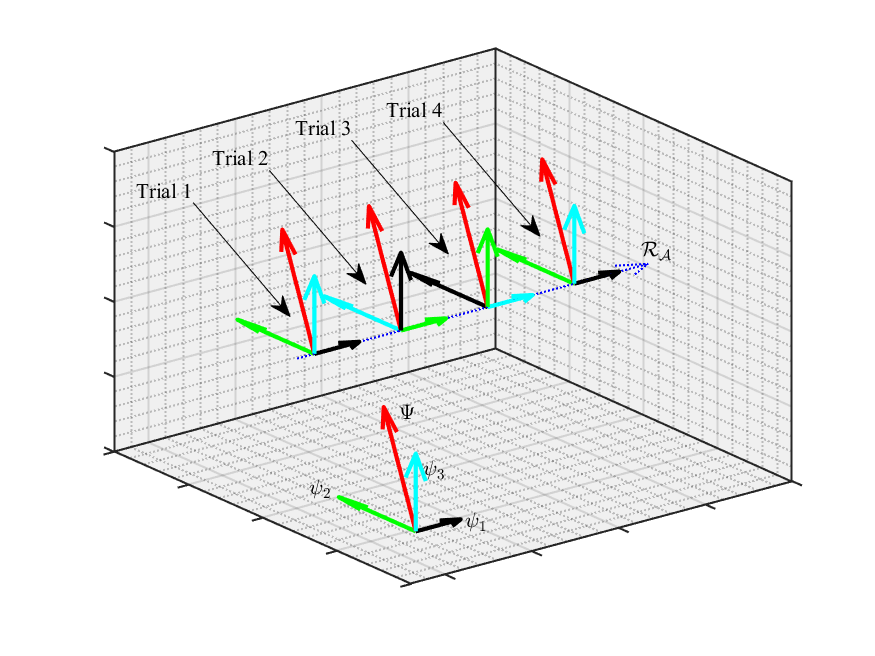}
        \caption{Trials of a random experiment, and the outcomes (states) which are observed, falling parallel to the reality axis $\mathcal{R_A}$ (blue, dotted line). Each outcome marked with different colours (black, green, cyan). Vector $\bm{\Psi
        }$ remains unchanged at each trial. Each rotation causes one (and only one) outcome to be expressed along the $\mathcal{R_A}$. The translation of $\bm{\Psi}$ to be ignored, and has been shown to distinguish between different trials.}
        \label{fig:trials_trails}
    \end{figure}
    
    Each trial rotates  $\bm{\Psi}$  such that one orthogonal component gets aligned to $\mathcal{R_A}$. Only this component is then expressed as the outcome of the experiment in that trial. This is highlighted in figure \ref{fig:trials_trails}. The act of rotation produces random orientations of the probable states. At least one, and only one outcome is always expressed out of the $n$ possible states. The problem now shits to a mechanical analogue: a random experiment corresponds to a rotation of the experiment vector $\Psi$, such that one of the outcomes $\ket{\psi_i}$ of the experiment gets expressed by coinciding to (or being recorded by) the reality axis $\mathcal{R_A}$. A rotational operator can be presented to the state vector $\ket{\bm{\Psi}}$, in order to perform this rotation. The rotation should be such that:
    \begin{itemize}
        \item A single basis state of  $\bm{\Psi}$  should be parallel to $\mathcal{R_A}$, and
        \item The state vector  $\bm{\Psi}$  should remain the same.
    \end{itemize}
    Both of these conditions can be met by rotating $\bm{\Psi}$ about its own axis as shown in figure 1. The rotational operator $\hat{\mathcal{R}}$ can be determined by shifting one orthogonal axis (say $\hat{z}$) along  $\bm{\Psi}$  and then performing rotation, for which the Rodriguez' Formula can be employed \cite{Rodrigues1840DesProduire, Cheng1989AnRotations}:
    \begin{gather}
        \text{For }\hat{n} = \frac{\bm{\Psi}}{|\Psi|},\\
        \hat{\mathcal{R}}(\theta) = \mathbb{I}\cos(\theta) + (1-\cos(\theta))(\hat{n}\otimes\hat{n}) + [\hat{n}]_{\times}\sin(\theta).
    \end{gather}
    Here $\mathbb{I}$ is the identity matrix and $[\hat{n}]_\times$ is the cross product matrix. The constraint on $\theta$, which should be that it is equal to $p\cdot 2\pi/n$, where $p \in [0,1,2 .... (n-1)]$, is not stressed upon, as will be shown later. At each ($j^{th}$) step, the outcome of the experiment can be determined as:
    \begin{gather}
        ^{j}\bm{\Psi} = \hat{\mathcal{R}}(\theta)\cdot\,\bm{\Psi}.
        \intertext{The expressed outcome can be found by determining the projection on the $\mathcal{R_A}$ axis:}
        \ket{\psi}_{expressed} = {\mathcal{R_A}}\cdot\,^{j+1}\bm{\Psi}.\\
        \ket{\phi}_{expressed} = \frac{\ket{\psi}_{expressed}}{\braket{\psi|\psi}_{expressed}^{1/2}}.
    \end{gather}
    At each step, the rotation operator rotates the state vector $\bm{\Psi}$ such that one outcome is expressed at each step. The choice of the angle can now be expressed as a random value between $0$ and $2\pi$.  It is reasonable to state that $\mathcal{R_A}$ must be a dyadic tensor of $n\times n$ dimensions. In figure 1, it corresponds to:
    \begin{equation}
        \hat{\mathcal{R_A}} = \begin{pmatrix}
            1&0&0\\0&0&0\\0&0&0
        \end{pmatrix}.
    \end{equation}

\begin{figure}
        \centering
        \includegraphics[width=1\linewidth]{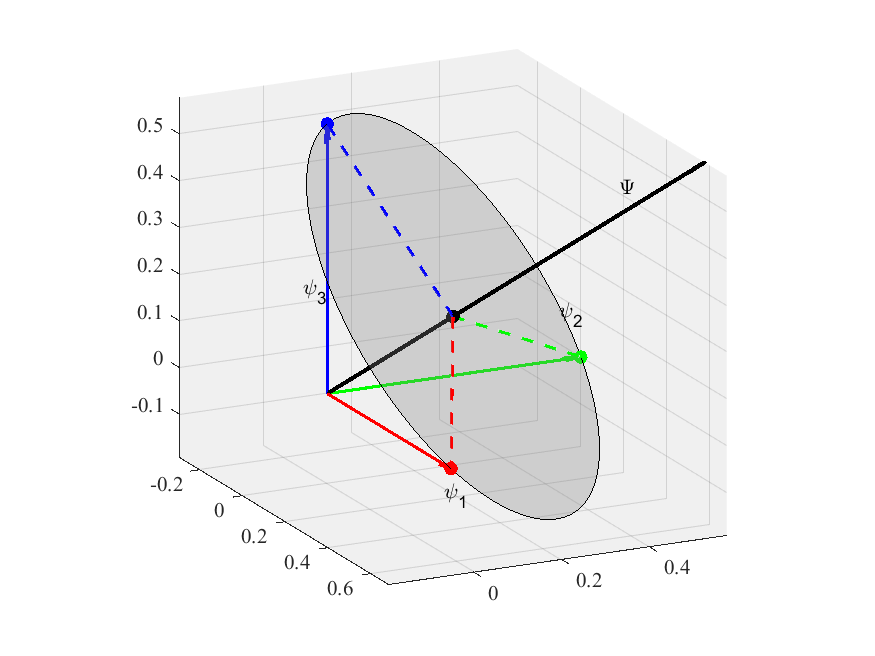}
        \caption{Circular modelling of the rotating vector state, with the projection of each outcome on the disc $\mathcal{D}$ shown as dotted line.}
        \label{fig:disc}
    \end{figure}

    In such a modelling, a disc containing projections of each possible  state $\ket{\psi_i}$ can be formed, with each basis presenting a projection $\bm{r}$ on the disc:
    \begin{equation}
        r = |\psi_i|\sin\left[ \cos^{-1}\left( \frac{1}{\sqrt{n}}\right) \right]
    \end{equation}
    where $\bm{r}$ is the radius of the disc. This disc, which has been termed in this work as $(\mathcal{D}_n)$, rotates with the state vector $\bm{\Psi}$. For the three-outcome case, the corresponding $\mathcal{D}_n$ is shown in figure \ref{fig:disc}.
    
    It is obvious that the disc is circular, with $P_i=1/n\,\forall\, i\in[1,n]$. In a general case, its structure is an ellipse, with the eccentricity proportional to $|n^{-n} -\Lambda_\sigma^0$|. The general elliptic structure, using a polar equation in its own plane, centred at the point $(P_1,P_2,P_3,....,P_n)$. For an $n-$dimensional outcome space, $\mathcal{D}_n$ can be framed by taking this point as the origin and using the plane-polar equation of origin-centred ellipse:
   \begin{gather}
       r(\theta) = \frac{b}{\sqrt{1-\epsilon^2\cos^2(\theta)}},
       \intertext{where $\epsilon$ is the eccentricity and $b$ is the semi-minor axis of the ellipse. The semi-major axis $a$ is related to these in the well-known relation,}
       a= r(0) = \frac{b}{\sqrt{1-\epsilon^2}}, \quad a>b.
   \end{gather}

    For the uniform case, it can be determined from equation (22) that 
    \begin{equation}
        a=\,b = \sqrt{P_i}\cdot\sin\left[\cos^{-1}\left(\frac{1}{\sqrt{n}}\right)\right], \quad i\in[1,n]. 
    \end{equation}
    In the uniformly distributed randomness case, the projection of each basis state $\ket{\psi_i}$ on the state vector $\bm{\Psi}$ is the theoretical probability of that outcome:
    \begin{equation}
        P_i = \ket{\psi_i}-\bm{r} = \braket{\psi_i|\Psi},\quad\because |\bm{\Psi}|\equiv 1.
    \end{equation}
    
\begin{figure}[!ht]
        \centering
        \includegraphics[width=1\linewidth]{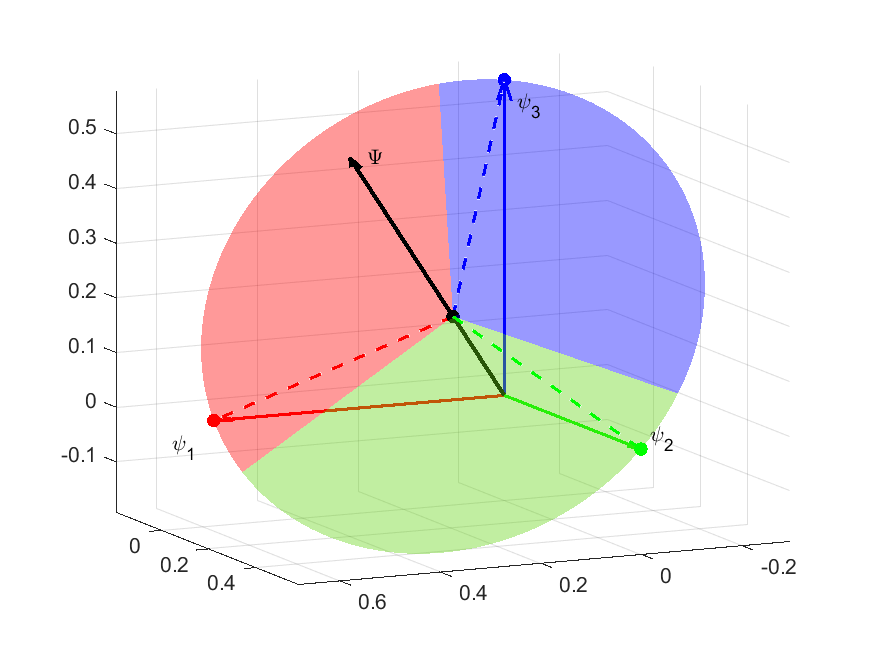}
        \caption{Sectors of $\mathcal{D}_n$ associated with the 3 outcome state. Each sector is associated with one outcome state vector $\ket{\psi}$, which is represented using the colour combinations.}
        \label{fig:enter-label2}
    \end{figure}

    The discreteness of the outcomes during each rotation implies a rigid rule: every rotation must be quantized to $\theta = 2\pi p/n$, where $p\in[0,n-1]$. To account for this descretisation, the disc can instead be divided into $n$ sectors, each having a central angle equal to $2\pi/n$. The placement of these sectors relative to the component $\ket{\psi_i}$ projections on the disc $(\bm{r}_i)$ is irrelevant and we show the case of 3 component vector state in figure \ref{fig:enter-label2}.\\
    
    Similar projected discs $\mathcal{D}_n$ can be framed for an $n$-outcome system, in the same way as shown in figure \ref{fig:enter-label2}. Here, it is highlighted that the model would be unaffected for 4 or higher dimensions\footnote{The argument here is -- it suffices to state that each component of such higher-dimensional vectors can be described to be independent of another using orthonormal bases, and the projections of each fill follow the same expressions as presented in equations (22) and (23). The inability of imagining higher dimensions can therefore be simply avoided.}. A ten-outcome state vector can therefore be represented by a disc with $10$ such projections $\bm{r}$, where
    $$ r_i = |\psi_i|\sin\left[\cos^{-1}\left( \frac{1}{\sqrt{10}}\right)\right],$$ and $$P_i = \bm{\Psi}-\bm{r} = \frac{1}{10}.$$

    The problem of random experiment repetition has now shifted to a mechanical problem of a disc rotation. We now describe the dependence of each outcome on the previous states of rotation in the next section. 

\section{Fluctuations in the initial conditions as a cause of randomness}
\begin{figure*}[!ht]
        \centering
        \includegraphics[width=0.8\linewidth]{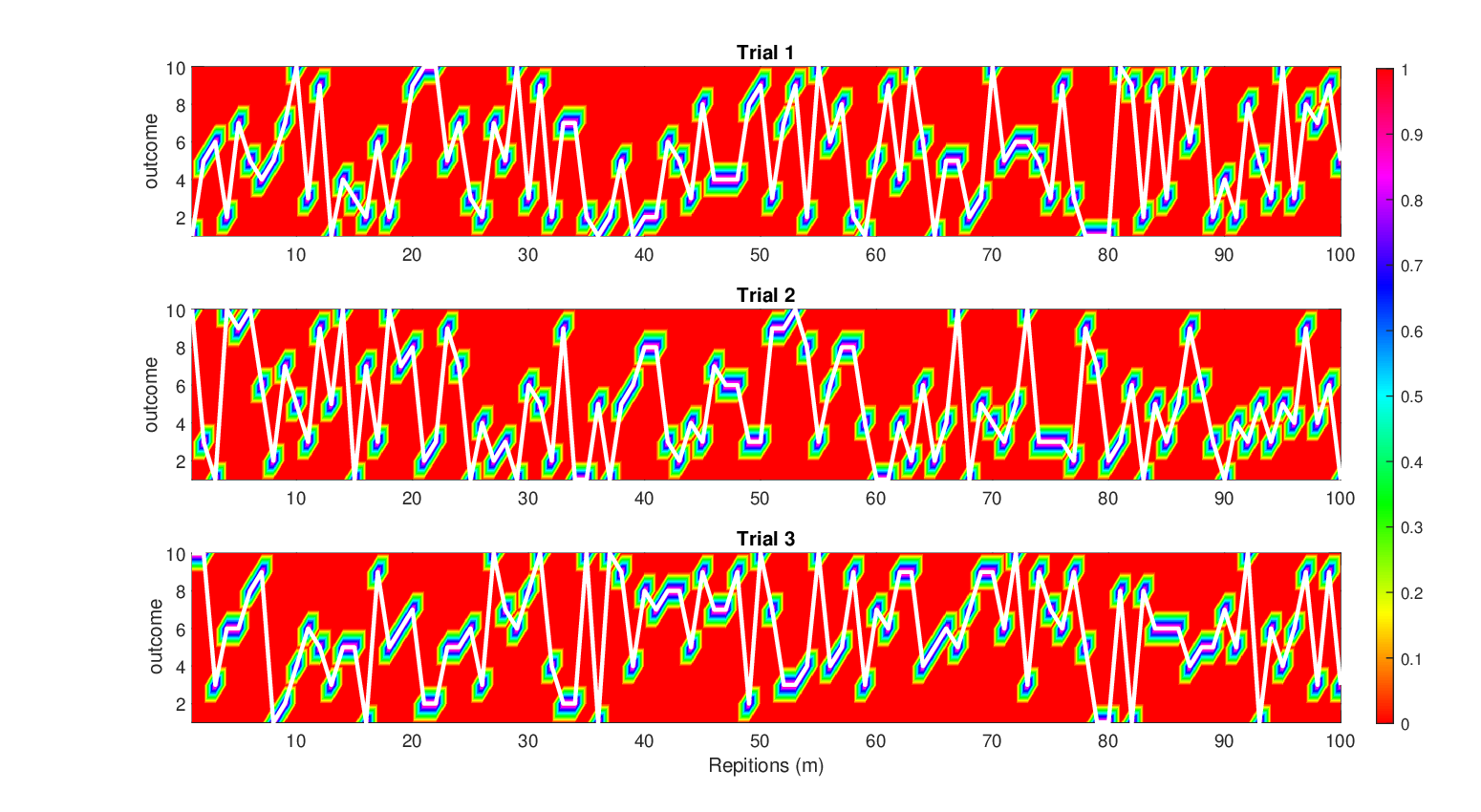}
        \caption{'Flow' of the reality axis across various possible outcomes, for $100$ repetitions, for $3$ runs. In the colour gradient, $1$ represents the true outcome, where-as $0$ represents the outcomes which do not occur for that repetition. It can be seen that each repetition presents a different angular rotation step, marked in this figure by the length of the white curve which connects each outcome.}
        \label{fig:flows}
    \end{figure*}

 Suppose when the experiment is performed for the first time, the outcome of the experiment becomes $^1\ket{\phi_0}$. This corresponds to an angular rotation of $\theta_0$ of the state vector $\bm{\Psi}$. Suppose this experiment is now repeated exactly as it was performed before, the angular rotation (which corresponds to the act of performing the experiment) should be identical. This should be obvious since an experiment repeated exactly as it was performed before, must yield the same outcome, as is expected from the homogeneity of fundamental physical laws. This means that the new angular displacement should be equal to $\theta_0$. Generalizing, this yields
    \begin{equation}
    \begin{rcases*}
        ^{j+1}\theta = \,^j\theta = \,^0\theta,\\
        ^{j+1}\ket{\psi} = \mathcal{R}(\theta_0)\cdot\ket{\bm{\Psi}}=\,^j\ket{\psi},
        \end{rcases*} \quad \forall j\in [1,M],
    \end{equation}
    from which follows that the same outcome must occur every time the experiment is repeated. In other words, each repetition should be, in the analogy of this rotating vector state, of the same angular shift, and hence the equation (22) should be followed\footnote{This idea of causality is not new; a coin-toss, when repeated with the exact force, exact initial position and exact external (environmental) influences, would yield the same result, no matter how many times the experiment is repeated.}. However, this is not true, in practice. as shown in figure \ref{fig:flows} for the case of an experiment with 10 possible, equally likely outcomes. 

    The \textit{jump}, or the angular rotation step at each trial does not remain constant. Assuming the experiment is repeated identically, with exact methods and approaches, the only change which must produce the fluctuation in the process is the initial states at the beginning of each trial, which change after each repetition. These changes are produced by the \textit{microscopic} deviations from the exact initial conditions, which include orientational, parametric and other conditions. These deviations are \textit{microscopic} since they are non-incorporable or negligible to the experimentalist or the observer. Incorporating such, otherwise neglected deviations, would make the physical problem either unsolvable or highly nonlinear, due to the presence of large number of unknowns, as is suggestive from a mechanical formulation of the act of performing the experiment.
      
    In the above discussion, it is both interesting and important to note that the presence and influence of these microscopic deviations is the root-cause of \textit{random} selection of possible outcomes. The inability to determine the outcome, lies not in absolute uncertainty, but in the inability to mechanically encounter such deviations. Hence, each successive rotation is separated by an addition of an unknown angular shift, which may be represented by some \textit{random} $\delta\theta_r$. 
    \begin{gather}
        ^{j+1}\theta = \,^j\theta + \delta\theta_r.\\
        ^{j+1}\ket{\phi} = \mathcal{R}({\,^0\theta})\cdot\left(\mathcal{R}({\delta\theta_r})\cdot\,^j\ket{\phi}\right).
    \end{gather}
    
\begin{figure}[!hb]
        \centering
        \includegraphics[width=0.9\linewidth]{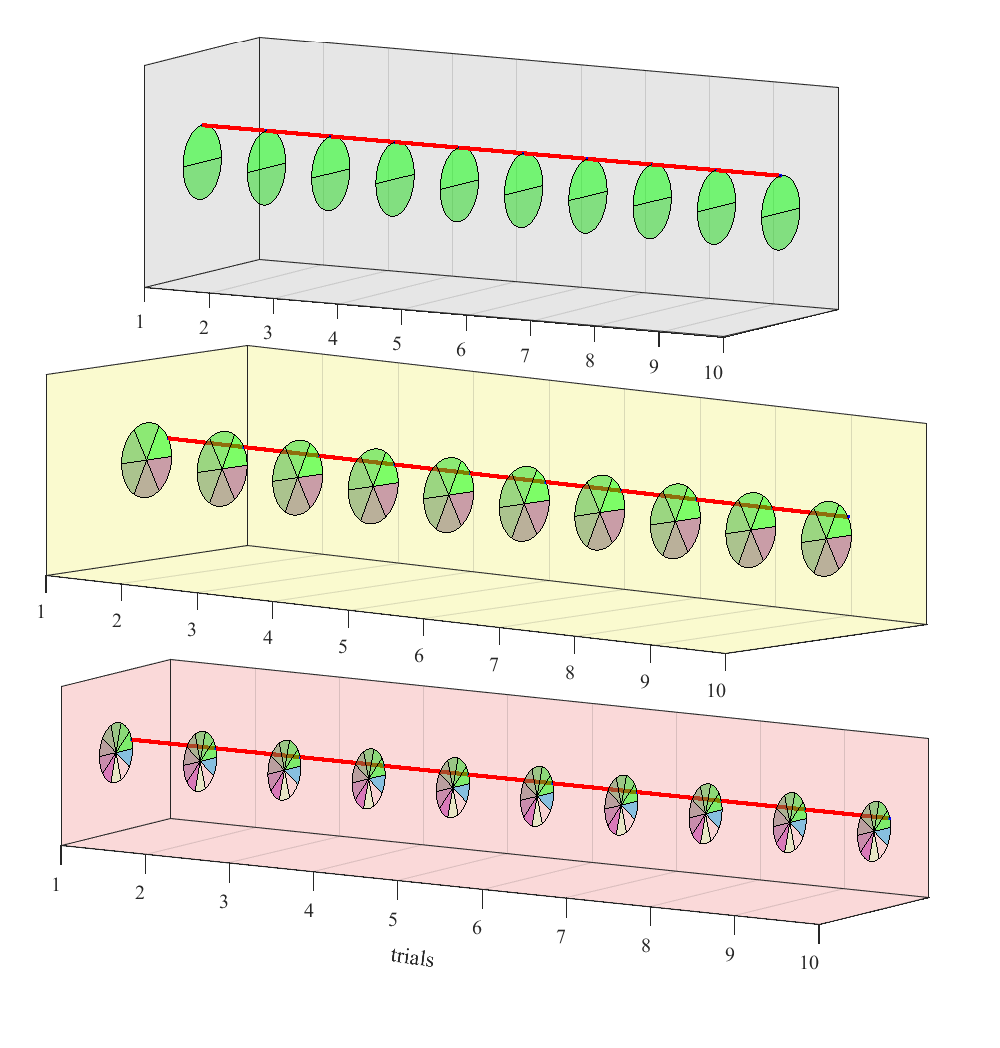}
        \caption{Random experiments with i.i.d. outcomes: (top) a coin toss, (middle) a die roll and (bottom) a $10$ outcome experiment. In each case, the influence of the microscopic deviations is not included, resulting in the same outcome occurring in each trial. Each experiment is represented by its own $\mathcal{D}_n$, with the coin toss having 2 sectors, the die roll having 6 sectors and the 10 outcome experiment having 10 sectors. The red curve in each case represents the reality axis $\mathcal{R_A}$.}
        \label{fig:unbiased_no_flux}
    \end{figure}

\begin{figure}[!ht]
        \centering
        \includegraphics[width=0.9\linewidth]{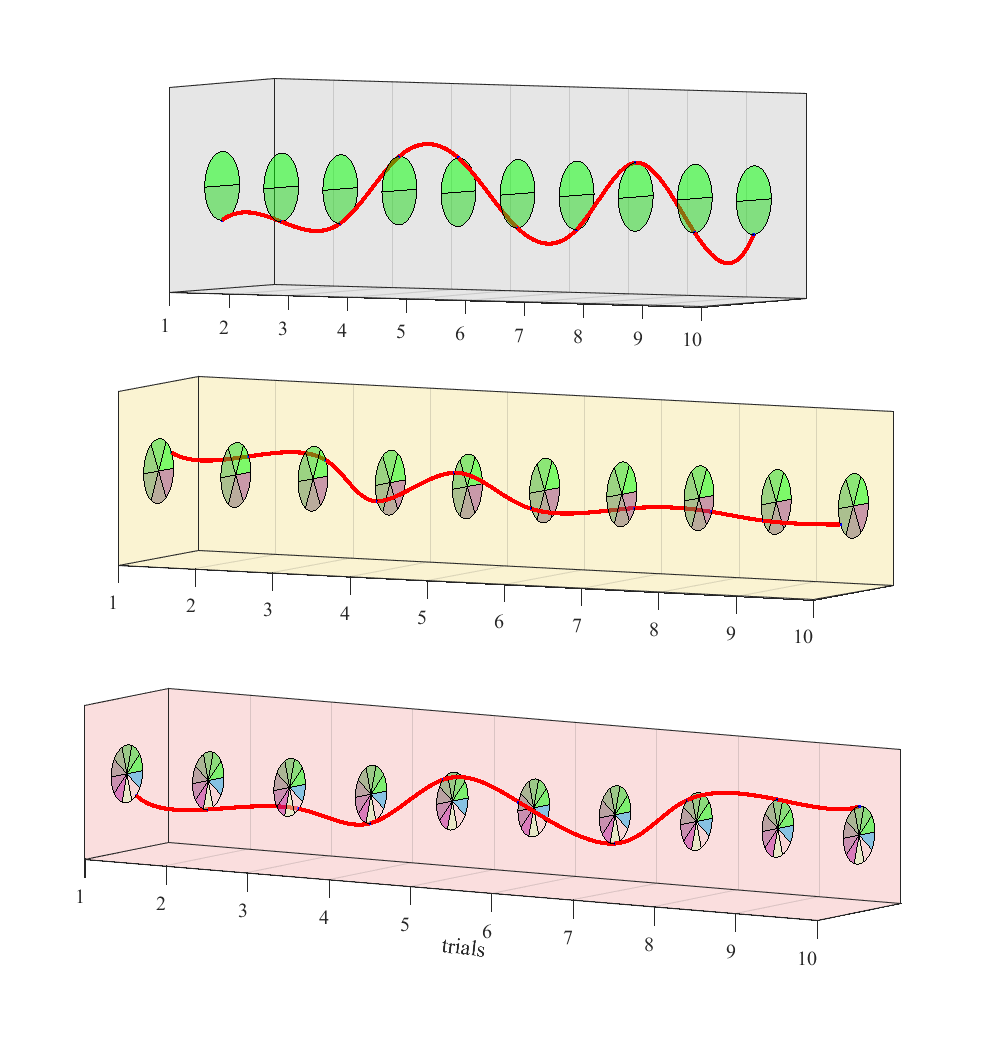}
        \caption{Random experiments with i.i.d. outcomes: (top) a coin toss, (middle) a die roll and (bottom) a $10$ outcome experiment. In each case, the influence of the microscopic deviations is included as a random angular shift of the initial state during the act of performing the experiment, resulting in the change in outcomes during each trial. Each experiment is represented by its own $\mathcal{D}_n$, with the coin toss having 2 sectors, the die roll having 6 sectors and the 10 outcome experiment having 10 sectors. The red curve in each case represents the reality axis $\mathcal{R_A}$. Figures represent the observer frame attached to $\mathcal{D}_n$ in each case, due to which the reality axis (and not $\mathcal{D}_n$ itself) rotates.}
        \label{fig:unbiased_flux}
    \end{figure}

\begin{figure}[!ht]
        \centering
        \includegraphics[width=1\linewidth]{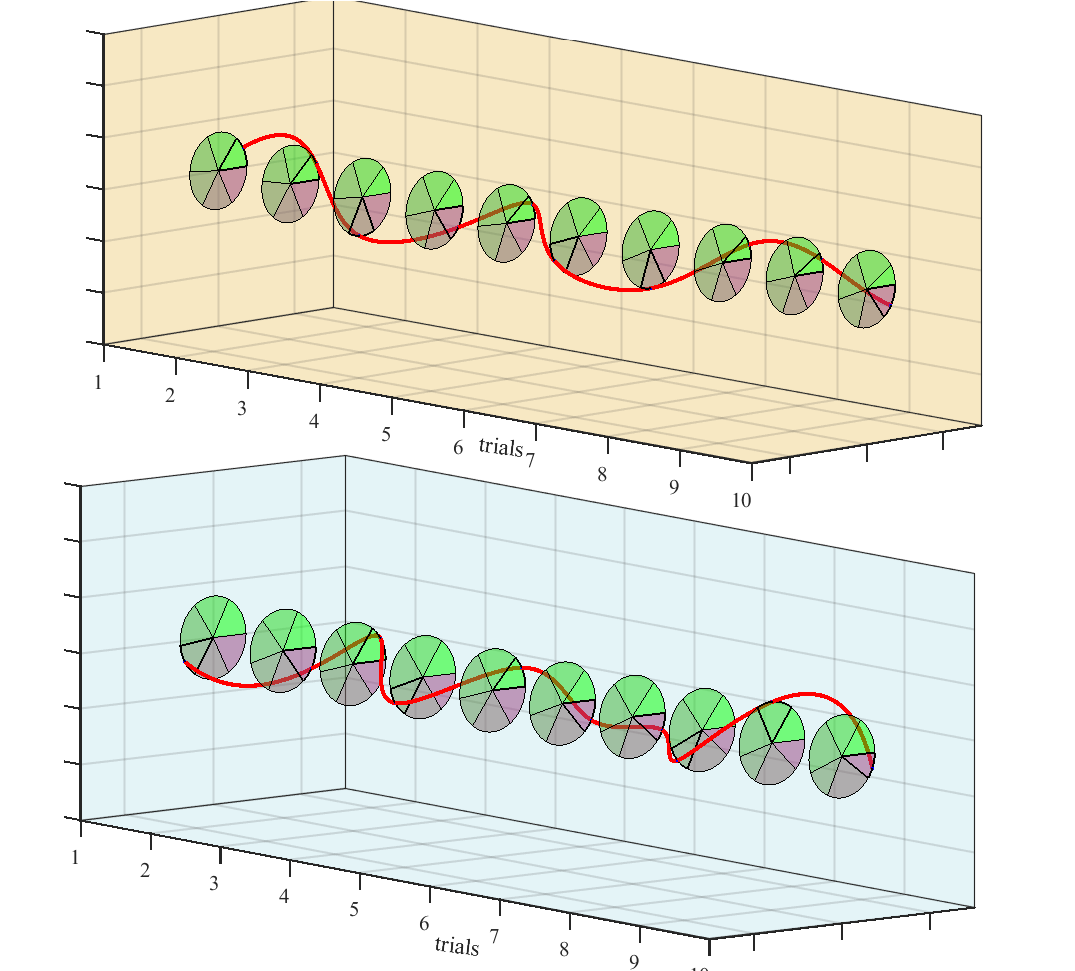}
        \caption{A random die roll with $6$ i.i.d. outcomes, each represented by a sector of $\mathcal{D}_6$. In the two runs (shown above and below), the reality axis follows random paths, relative to the $\mathcal{D}_6$ frame. The influence of the microscopic deviations in the initial conditions are therefore the cause of $randomness$ in an experiment in repetition.}
        \label{fig:randomens}
    \end{figure}

    From the above discussions, one implies that the magnitude of angular displacements change at the $(j+1)^{th}$ trial because of addition of some angular shifts to the expressed outcome $^j\ket{\phi_i}$, due to these \textit{microscopic} deviations. Each angular shift causes a different outcome state to be aligned to the reality axis $\mathcal{R_A}$. Keeping our observation frame attached to $\mathcal{D}$, one can show how the outcomes of an experiment in repetition are recorded. This is shown for the coin-toss, dice roll, and a $10$ outcome random experiment in figures \ref{fig:unbiased_no_flux} and \ref{fig:unbiased_flux}. Due to the fluctuation in the initial state during each repetition of the experiment, a new outcome is produced, which when ignored, causes the same outcome to be produced repeatedly. The influence of these microscopic fluctuations are not considered during each repetitions, and this gives rise to random outcomes. Figure \ref{fig:randomens} shows the randomness in the trajectory of the reality axis during two runs of a die roll, 10 times in each run.


    At each trial, these fluctuations cause $\mathcal{D}_n$ to rotate by an unpredictable angle, unpredictable due to their influence being neglected. This causes $\mathcal{D}_n$ to rotate such one sector touches the reality axis $\mathcal{R_A}$. This causes the associated outcome state $\ket{\psi}$ to be recorded and therefore expressed as the true outcome. \textit{Randomness}, therefore, is produced only when an experiment is repeated at least once, causing the fluctuations to play a role. This readily implies that an experiment with i.i.d. outcomes is not random, unless repeated. Such an experiment, when performed only once, produces outcomes which are exactly predictable, provided that the exact initial conditions are incorporated in the physical mechanism of the experiment.

    It can be seen that such an influence of the microscopic deviations in the initial conditions of an experiment play a role in the rotation of $\mathcal{D}_n$, which itself is geometrically orthogonal to the state vectors $\psi_i$. This implies that the probability $P_i$ of any outcome, which is simply the likelihood of its occurrence, remains the same. Probability, therefore, does not play a role in the randomness of the outcomes of an experiment during its repetition. It can be seen (from figure \ref{fig:unbiased_no_flux}) that even upon ignoring the fluctuations in the initial conditions, the probability of each outcome remains the same. The empirical repetition of a random experiment and the probability of its outcomes are therefore two independent physical entities, not influencing each other during the dynamics of the experiment.

    The \textit{random} choice of the fluctuation angle $\delta\theta_r$ is, once again, due to the incapacity of the experimentalist or the observer to account for the exact change in the initial conditions of the experiment. However, this random choice of $\delta\theta_r$ implies the true origin of the convergence of $\Lambda_\sigma$ towards $\Lambda_\sigma^0$ following the convergence of $\mathcal{L}_i$ to $P_i$, as is displayed by LLN. In a kinematical set-up, suppose that a jumper takes jump of length $S$, the value of which is unknown. The line in front of the jumper is divided into $n$ equal parts. This is shown in the following diagram:
    \begin{center}
\begin{tikzpicture}[scale=1, every node/.style={font=\small}]

% Line divided into 3 segments
\draw[thick] (3,0) -- (6,0);
\foreach \x in {0,1,2,3} {
  \draw[thick] (3+\x,0.1) -- (3+\x,-0.1);
}
\node at (3.5,-0.4) {n=1};
\node at (4.5,-0.4) {n=2};
\node at (5.5,-0.4) {n=3};

% Stick figure person
\draw[thick] (0,0) -- (0,1.2);             % body
\draw[thick] (-0.3,1.5) arc (180:540:0.3); % head
\draw[thick] (0,1.2) -- (-0.5,0.8);        % left arm
\draw[thick] (0,1.2) -- (0.5,0.8);         % right arm
\draw[thick] (0,0) -- (-0.3,-0.6);         % left leg
\draw[thick] (0,0) -- (0.3,-0.6);          % right leg

% Curved arrow (jump to Sector 2)
\draw[->, thick, decorate, decoration={snake, amplitude=0.5mm, segment length=3mm}]
  (0.3,1) to[out=20, in=160] (4.5,0.3);

% Optional: label the person
\node at (0,-1) {jumper};
\end{tikzpicture}\end{center}
    Now, suppose the jumper jumps in this region for a total of $m$ times. Let the total number of times that it jumps in a segment $i$ is $m_i$, and the total number of times it jumps on another segment $k$ is $m_k$. It is an obvious kinematical conclusion that 
    $$|m_k - m_i| \propto |k-i|.$$
    This means that segments closer to each other will have landing frequencies closer to each other. This analogy offers a heuristic motivation for the angular proximity rule.
    
    Suppose the experiment $\bm{\Psi}$ occurs with $n$ possible outcomes. At the end of $j^{th}$ trial, the empirical probability of some outcome $\ket{\phi_i}$ is $\mathcal{L}_i = m_i/j$. This means that during the $j$ angular rotations of $\mathcal{D}_n$, $\mathcal{R_A}$ has met the $i^t{h}$ sector for $m_i$ times. Simultaneously, another outcome $\ket{\phi_k}$ gains number of outcomes $m_k$ such that $\mathcal{L}_k=m_k/j$. It is then an obvious kinematical conclusion that
    \begin{gather}\label{ssoolln}
        |m_k - m_i| \propto |\Delta(\theta_k,\theta_i)|,
        \Rightarrow  |m_k - m_i| = c |\Delta(\theta_k,\theta_i)|,
        \intertext{where $\Delta(\theta_k,\theta_i)$ represents the distance between the sectors $i$ and $k$ and $c$ is some constant. The above equation simply states that sectors (angular positions) closer to each other must have closer frequencies. Now, using some manipulations,}
       \frac{|m_k - m_i|}{m_k} = c \frac{|\Delta(\theta_k,\theta_i)|}{m_k}.
       \intertext{Suppose when the experiment repeats such that $m\to\infty$, at-least one outcome frequency (say $m_k$) proceeds towards infinity. This is obvious since at-least one outcome should occur for infinite time, when $m\to\infty$. We then get:}
       \lim_{m_k\to\infty}\left(1-\frac{m_i}{m_k}\right) = \frac{|\Delta(\theta_k,\theta_i)|}{m_k}.
       \intertext{The r.h.s. of the above equation will naturally become zero, since $|\Delta(\theta_k,\theta_i)|$ is fixed due to the structure of $\mathcal{D}_n$. Therefore,  }
        \lim_{m_k\to\infty} 1-\frac{m_i}{m_k} = 0 \implies m_i = m_k. 
    \end{gather}
    This means that as the number of trials increase towards infinity, the frequency of each outcome becomes equal to the other. This causes $\mathcal{L}_i$ to converge to equal values for all $i,j\in[1,n]$. As a result, both $\mathcal{L}_i$ and $\Lambda_\sigma$ converge to $P_i$ and $\Lambda_\sigma^0$ respectively. This evidently shows how the Law of Large Numbers comes into play. The LLN convergence is therefore an implication of the dynamical mechanics of the random experiment, incorporating the microscopic deviations in the initial conditions into the statistico-mechanical modelling shown in this section. 

    The consequential dynamical evolution of both individual outcome frequencies $m_i$ and the $\Lambda-$entropy have been derived in the previous work \cite{Lobo2025OnTheorem}, and have been represented in equations (\ref{dmkdn}) and (\ref{dlambdadm}). In both of these expressions, it can be seen that:
    \begin{itemize}
        \item The growth rate of one outcome frequency $m_k$ depends on the growth rate of other outcome frequencies $m_i$.
        \item The evolution of $\Lambda_\sigma$ depends on the evolution of empirical probabilities $\mathcal{L}_i$, relative to each other. 
    \end{itemize}
     This raises some interesting questions:
    \begin{enumerate}
        \item Does the outcome of the $j^{th}$ trial of a random experiment affect the $(j+1)^{th}$ outcome?
        \item If so, how?
    \end{enumerate}
     We address these questions in the next section.
     
     \section{Entanglement between successive outcomes}
    
As discussed above, in the results of the analyses shown in section 2, specifically in the study of the evolution of $m_i$ and $\Lambda_\sigma$ with $m$, it is seen that these evolutions depend on some feedback produced by the empirical frequency and probabilities of each outcomes. There therefore exists some coupling between each outcomes of the $(j+1)^{th}$ trial with the outcomes of the $j^{th}$ trial. From equation (\ref{dmkdn}) it is clear that a feedback is presented to the evolution of the outcome frequencies $(m_i)$, which is based on the memory of the existing empirical probabilities $\mathcal{L}_{i,k}$. This means that nature prefers the $i^{th}$ outcome if its empirical probability $\mathcal{L}_i$ is lesser than that of other outcomes. In other words, outcomes with lesser frequencies $m_i$ are preferred. Also, the convergence rate of $\Lambda_\sigma$ is decreased by the increase in number of possible outcomes $(n)$. Based on these qualities, a feedback mechanism is proposed, as follows:

\subsection*{Novelty and memory bias in random experiments}
The act of performing, and repeating the random experiment, in the model described above, is similar to a spinning roulette disc. Each trial causes $\mathcal{D_f}$ to spin, and the expressed outcome is decided by the association of the reality axis $\mathcal{R_A}$ with the sector of the outcome, as in the case of a roulette disc. This association is that $\mathcal{R_A}$ exhibits that outcome $\ket{\phi_i}$, who's sector meets (associates) with the $\mathcal{R_A}$. Now, in the initial set-up, each sector has equal central angle $(\theta_i)$ area $(a_i)$ and arc-length $(S_i)$, given by:
     \begin{equation}
     \begin{rcases*}
         \theta_i = 2\pi P_i,\\ S_i= \bm{r}_i\theta_i = 2\pi\bm{r}_iP_i,\\ a_i=r_i^2\frac{\theta_i}{2} =a_i=\pi P_ir_i^2 .
     \end{rcases*}    
     \end{equation}
     As the experiment proceeds, in order to show the modification in the \textit{preference} of outcomes, such that $\mathcal{L}_i$ tends towards $P_i$ as $j$ increases, a bias must be introduced in the sectors associated with each outcome. This implies that, even though each outcome $\ket{\phi_i}$ still remains equally likely (since either projections of $\ket{\psi_i}$ remain unchanged), nature tends to \textit{prefer} such outcomes which have lesser $\mathcal{L}_i$. This preference has been discussed already, as shown by Lobo and Arumugam \cite{Lobo2025OnTheorem}. The bias can be presented as a factor which is multiplied to each sector angle: 
     \begin{gather}
         \theta_i\rightarrow\begin{cases}
             \alpha\theta_i,  & \text{$\ket{\phi_i}$ is expressed},\\
             \beta_{ki}\theta_k, & \text{$\ket{\phi_k}$ is not expressed}.
         \end{cases}
         \intertext{Here, $\alpha$ and $\beta_i$ are parameters which express habituation decay and novelty growth, and are related to the experiment as follows:}
         \alpha=1-\frac{1}{n^q},\quad \beta_k =\frac{\theta_i(1-\alpha)}{\theta_k(n-1)} + 1.
     \end{gather}
    The choice of $\alpha$ follows from the observation that the rate of convergence of $\mathcal{L}_i$ to $P_i$ is inversely proportional to number of possible outcomes $n$ \cite{Lobo2025OnTheorem}. The factor $\beta$ then simply distributes the angular reduction from $\theta_i$ equally among all non-expressed $\theta_k$ sector angles. As is clear, $\alpha<1$ and $\beta>1$ in small limits of $n$, whereas both these parameters converge to $1$ in the limit $n\to\infty$.\\ 
    
    This model creates a biasing in a plane orthogonal to the actual expressions of true probability of each outcome; the bias arising purely out of fluctuations in values of $|\Lambda_\sigma-\Lambda_\sigma^0|$. The orthogonality shows that even though each outcome is still equally probable or likely, they are not equally preferred by nature. This preference depends on how frequently an outcome is expressed by an experiment in repetition. 

\begin{figure}[!ht]
        \centering
        \includegraphics[width=0.8\linewidth]{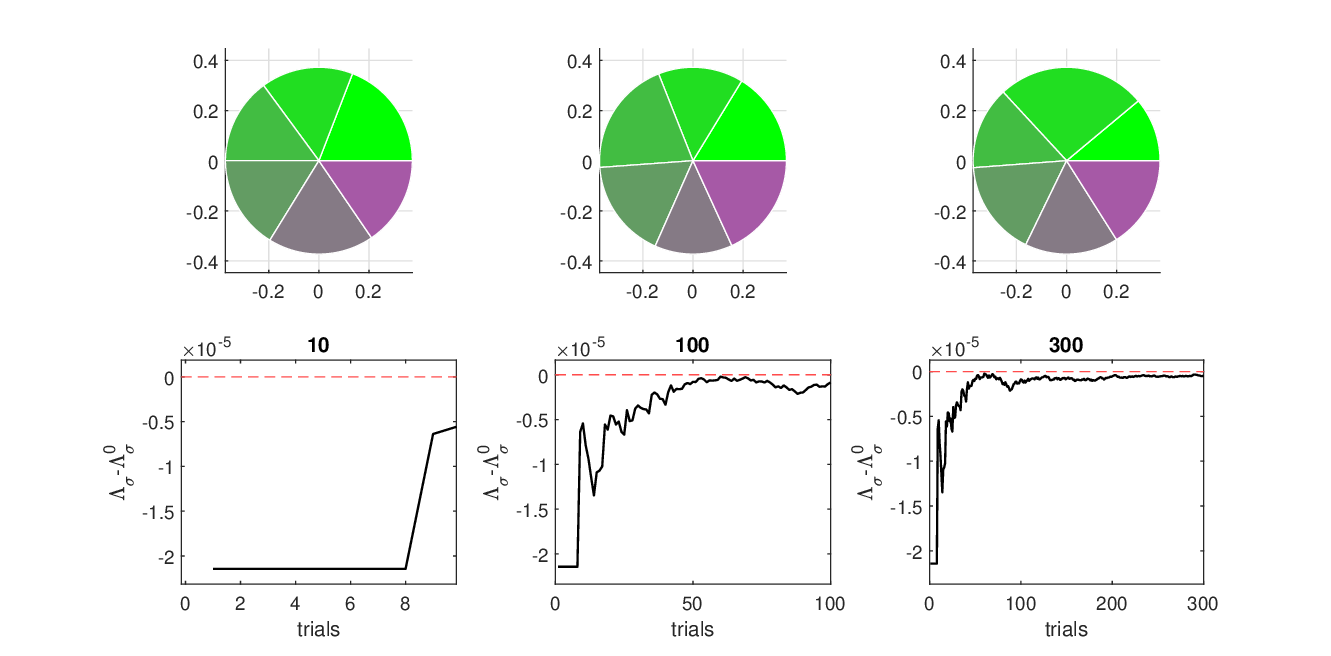}
        \caption{(Top) Evolution of sectors of $\mathcal{D}_6$, taking feedback order $q=1$. Simulation run for (left) 10, (middle) 100 and (right) 300 repetitions. (Below) Evolution and convergence of $\Lambda_\sigma$ entropy. }\label{fig:finalima0}
        \includegraphics[width=0.8\linewidth]{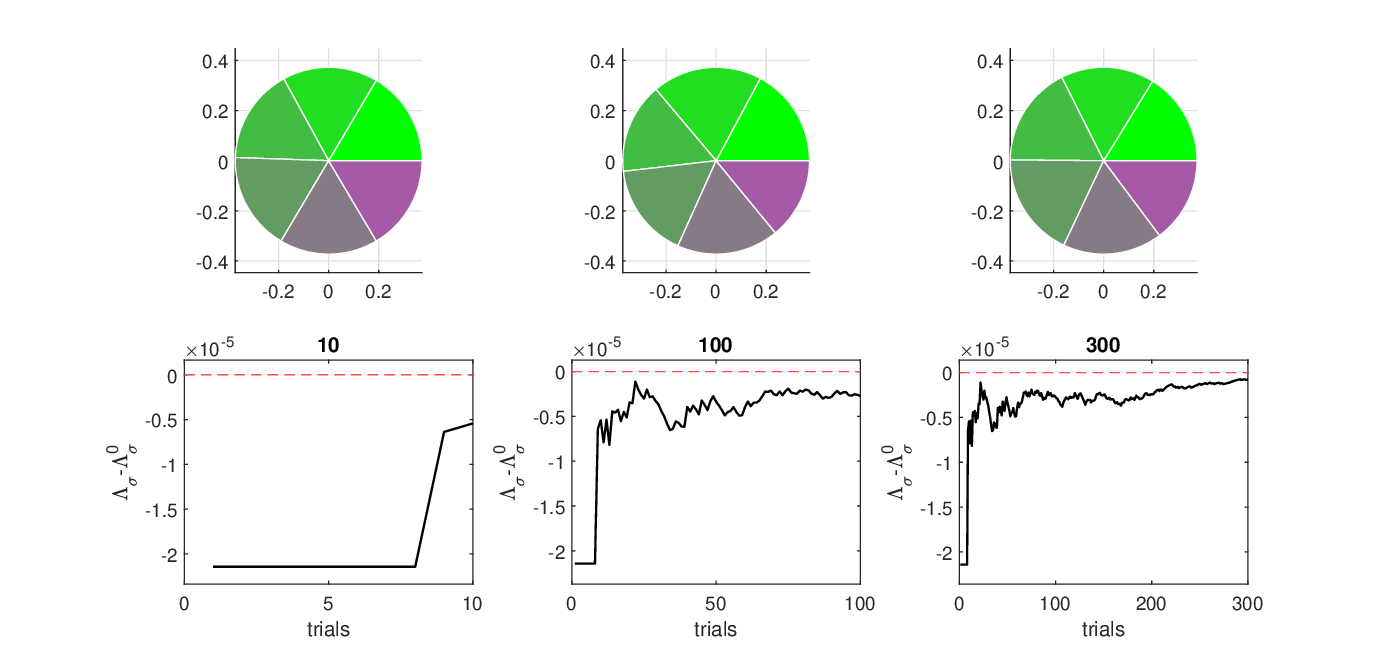}
        \caption{(Top) Evolution of sectors of $\mathcal{D}_6$, taking feedback order $q=2$. Simulation run for (left) 10, (middle) 100 and (right) 300 repetitions. (Below) Evolution and convergence of $\Lambda_\sigma$ entropy. }
        \label{fig:finalima1}
    \end{figure}

\begin{figure}[!ht]
        \centering
         \includegraphics[width=0.8\linewidth]{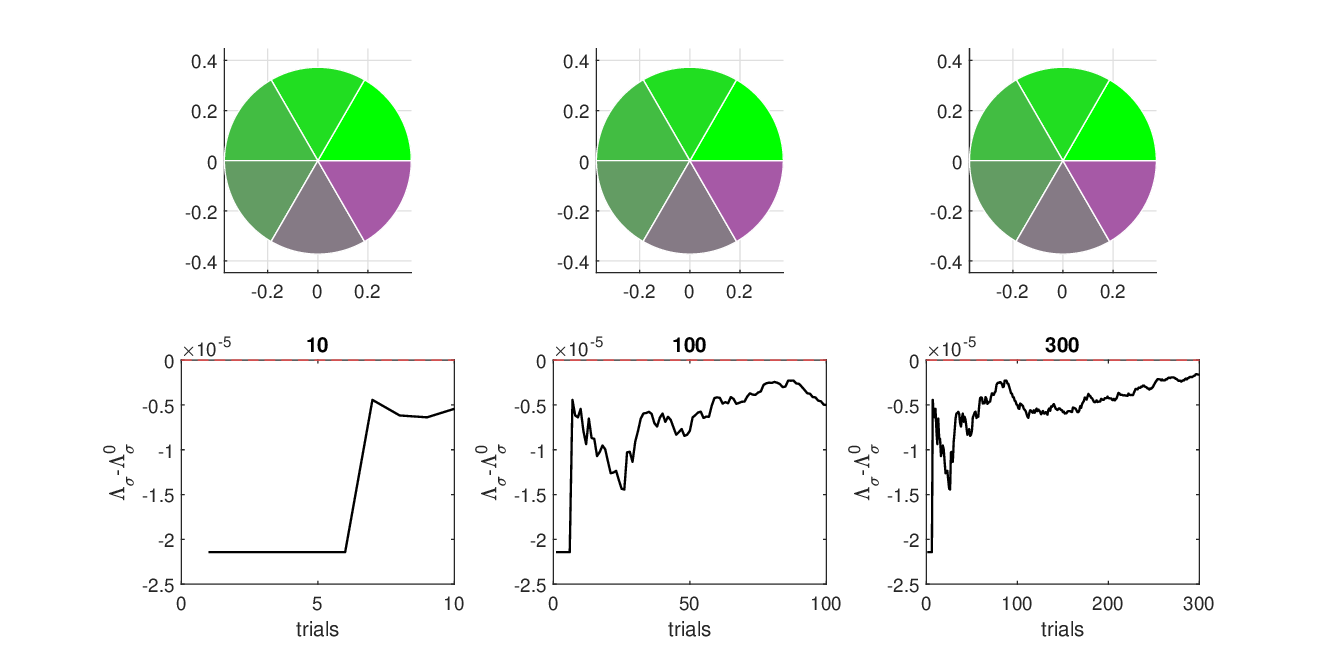}
        \caption{(Top) Evolution of sectors of $\mathcal{D}_6$, taking feedback order $q=5$. Simulation run for (left) 10, (middle) 100 and (right) 300 repetitions. (Below) Evolution and convergence of $\Lambda_\sigma$ entropy. }\label{fig:finalima1.5}
         \includegraphics[width=0.8\linewidth]{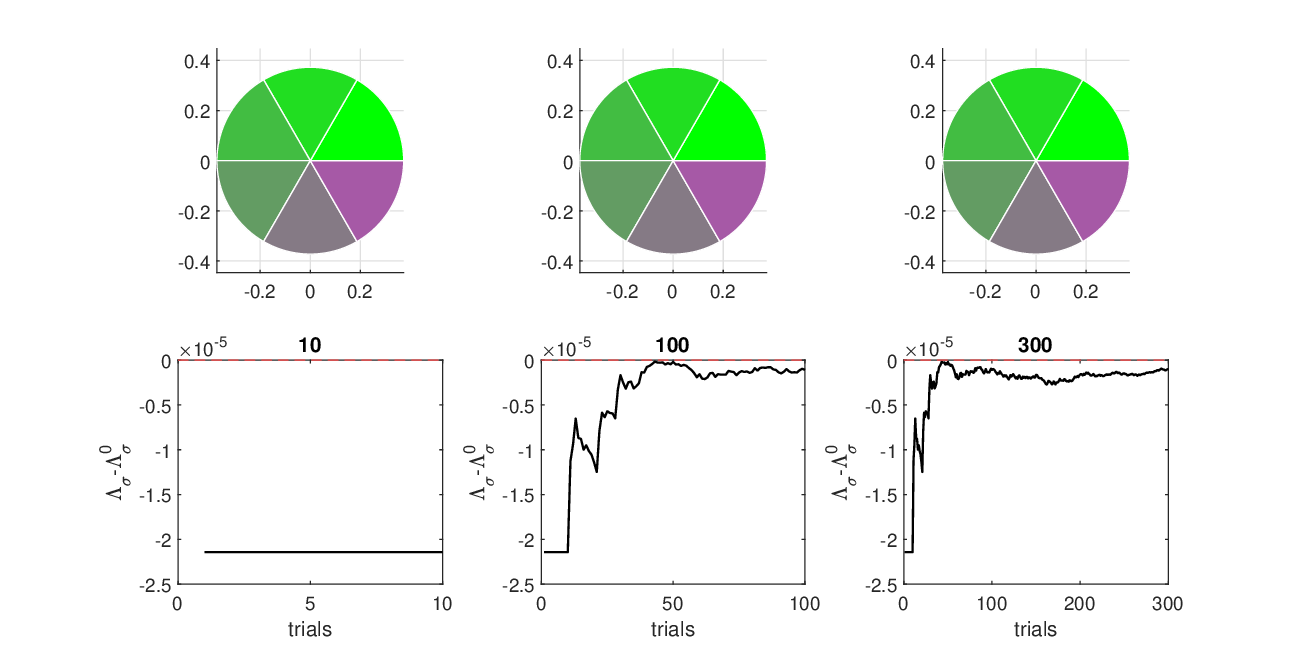}
        \caption{(Top) Evolution of sectors of $\mathcal{D}_6$, taking feedback order $q=\infty$. Simulation run for (left) 10, (middle) 100 and (right) 300 repetitions. (Below) Evolution and convergence of $\Lambda_\sigma$ entropy. }
        \label{fig:finalima2}
    \end{figure}

    The choice of the order of the feedback, $q$, can be determined by the rate of convergence of $\Lambda_\sigma$, and can be equal to $1,2,3 ..$. Higher orders of this feedback cause a slower convergence of $\Lambda_\sigma$. This is shown in figures \ref{fig:finalima0}, \ref{fig:finalima1}, \ref{fig:finalima1.5} and \ref{fig:finalima2}. 

     The presented mechanism provides a memory-based feedback to the evolution of the random experiment. It presents an idea of \textit{preference} in the system. \textit{Preference} is an entirely orthogonal phenomena to probability. Equally probable outcomes can have different preferences in the system, and equally preferred outcomes may bear different probabilities. The presented bias mechanism makes those outcome states $\ket{\phi}_i$ to be more preferred during an outcome, which have lesser empirical frequency. Although each outcome bears equal probabilities throughout the trials, they differ continuously in their preferences. This preference bears an order $q$, which determines how the system converges into pure empirical randomness. We now conclude our study in the next section.
\section{Conclusion}
In this work, I have proposed a dynamical framework for understanding the emergence of empirical convergence in repeated experiments. Departing from the traditional treatment of randomness as an assumed statistical condition, the model developed here constructs a rotation-based evolution of a state vector in outcome space, combined with feedback-sensitive biasing mechanisms and a projection process that determines realized outcomes.

The framework utilises the $\Lambda-$entropy \cite{Lobo2025OnTheorem} to quantify empirical randomness, which evolves over trials and serves as both a diagnostic of convergence and a parameter influencing the system's dynamics. The statistico-mechanical evolution of the state, which is influenced by the mechanically neglected microscopic deviations, orthogonal to the theoretical probability measurements, along with sectoral biasing terms $(\alpha$ and $\beta)$, guides the system toward statistical equilibrium, thereby reproducing the Law of Large Numbers as an emergent outcome of the process rather than a prior assumption.

The model recovers classical convergence in the large-trial limit and additionally captures transient effects and subtle entanglement between successive outcomes—features often absent from conventional stochastic descriptions. The geometric treatment of outcome space, along with the incorporation of memory through bias and projection, offers a new way to interpret convergence as a structured, self-correcting process.

While the current work focuses on uniformly distributed random experiments, the formulation is sufficiently general to accommodate extensions to biased distributions, adaptive feedback processes, and possibly quantum or biological systems where memory and recurrence play central roles. Further simulation and experimental validation would be important next steps toward assessing the broader relevance of this framework.

Ultimately, I hope this study contributes to the broader effort to understand randomness not only as a statistical abstraction but as a process with internal structure and causally evolving dynamics. The ideas developed here are intended as a foundation for further investigation into the mechanisms underlying empirical regularity in complex systems.\vskip6pt

\enlargethispage{20pt}

\paragraph*{Acknowledgement:}I thank the St Joseph's Research Institute, for providing the computational facilities, for which I am also very much grateful to Fr. Dr. Roshan Castelino, SJ.
\paragraph*{Data availability statement:}
The numerical codes (MATLAB) used to simulate the experiments and associated figures presented in this work are attached as supplementary material. There is no data associated with this work to be shown.
\paragraph*{Conflict of Interest:} No conflict of interest to declare. 
\paragraph*{Funding Statement:} No funding or research grant to declare.

%%%%%%%%%% Insert bibliography here %%%%%%%%%%%%%%

\vskip2pc
\bibliography{references}
\bibliographystyle{ieeetr}

\end{document}